\documentstyle[epsfig,graphics]{lamuphys}

\begin{document}

\title{Baryon Structure and the \\ Chiral Symmetry of QCD
}
\author{Leonid Ya. Glozman}
\institute{Institut f\"ur Theoretische Physik,
Universit\"at Graz,\\ A-8010 Graz, Austria}
\maketitle

\begin{abstract}
Beyond the spontaneous chiral
symmetry  breaking scale light and strange baryons
should be considered as  systems
of three constituent quarks with an effective confining interaction and a
chiral interaction that is mediated by the octet of Goldstone bosons
(pseudoscalar mesons) between the constituent quarks.
\end{abstract}

\section{Introduction}
Our aim in physics is not only to calculate some observable and get a
correct number but mainly to understand a physical picture responsible
for the given phenomenon. It  very often happens that a theory formulated
in terms of fundamental degrees of freedom cannot answer such a question
since it becomes overcomplicated at the related scale. Thus a main task in this
case is to select those degrees of freedom which are indeed essential.
For instance, the fundamental degrees of freedom in crystals are ions
in the lattice, electrons and the electromagnetic field. Nevertheless, in order
to understand electric conductivity, heat capacity, etc. we instead work
with  "heavy electrons" with dynamical mass, phonons and their interaction.
In this case a complicated electromagnetic interaction of the electrons with
the ions in the lattice is "hidden" in the dynamical mass of the electron
and the
interactions among ions in the lattice are eventually responsible  for the
 collective excitations of the lattice - phonons,
which are Goldstone bosons of the spontaneously broken translational
invariance in the lattice of ions.
As a result, the theory becomes rather
simple - only the electron and phonon degrees of freedom and their interactions
are essential for all the  properties of crystals mentioned above.
\footnotetext{Lecture given at the 35. Universit\"atswochen f\"ur
Kern- und Teilchenphysik, Schladming, Austria, March 1996
(Perturbative and Nonperturbative Aspects of Quantum Field Theory, ed. by
H. Latal and W. Schweiger, Springer 1996).}

Quite a similar situation takes place in QCD. One hopes that sooner or later
one can solve the full nonquenched QCD on the lattice and get the
correct nucleon and
pion mass in terms of underlying degrees of freedom: current quarks and
gluon fields. However, QCD at the scale of 1 GeV becomes too complicated,
and hence it is rather difficult to say in this case what kind of
physics, inherent in
QCD, is relevant to the nucleon mass and its low-energy properties. In this
lecture I will try to answer this question. I will show that it is the
spontaneous breaking of chiral
symmetry  which is the most important QCD phenomenon
in this case, and that beyond the scale of spontaneous breaking of chiral
symmetry light and strange baryons can be viewed as systems of three
constituent quarks which interact by the exchange of Goldstone bosons
(pseudoscaler mesons) and are subject to confinement.

It is well known that at low temperature and density the approximate
$SU(3)_{\rm L} \times SU(3)_{\rm R}$
chiral symmetry of QCD is realized in the hidden Nambu-Goldstone mode.
The hidden mode of chiral symmetry
is revealed by the existence of the octet of  pseudoscalar
mesons of low mass which represent the associated approximate
Goldstone bosons. The $\eta'$ (the $SU(3)$-singlet) decouples from the original
nonet because of the $U(1)_{\rm A}$ anomaly \cite{WE,THO}.
Another consequence of the spontaneous breaking of the approximate
chiral symmetry of QCD is that the valence quarks acquire
their dynamical or constituent mass
\cite{WEIN,MAG,SHU,DIP}
through their interactions with the collective excitations of
the QCD vacuum -
the quark-antiquark excitations and the instantons.

We have recently suggested \cite{GLO1,GLO2}
that beyond the scale of spontaneous breaking of chiral symmetry
 a baryon should be considered
as a system of three constituent quarks
with an effective quark-quark interaction that is formed
by a central confining part
and a chiral
interaction  mediated by the octet of pseudoscalar mesons
between the constituent quarks.

The simplest representation of the most important
component of the interaction of the constituent quarks  mediated
by the octet of pseudoscalar bosons in the $SU(3)_{\rm F}$
invariant limit is

\begin{equation} H_\chi\sim -\sum_{i<j}V(\vec r_{ij})
\vec \lambda^{\rm F}_i \cdot \vec \lambda^{\rm F}_j\,
\vec
\sigma_i \cdot \vec \sigma_j.\label{1.1} \end{equation}
Here  $\{\vec \lambda^{\rm F}_i\}$ represents the flavor $SU(3)$ Gell-Mann
matrices and the
 sums  run over the constituent quarks.

Because of the flavor dependent factor $\vec\lambda_i^{\rm F} \cdot
\vec\lambda_j^{\rm F}$ the chiral boson exchange interaction (\ref{1.1})
will lead to orderings of the positive and negative parity
states in the baryon spectra, which agree with the observed
ones in all sectors. For the spectrum of the nucleon
and $\Delta$ the strength of the
chiral interaction
between the constituent quarks is sufficient to shift
the lowest positive parity states
in the $N$=2 band (the ${\rm N}(1440)$ and $\Delta(1600)$)
below the negative parity states
in the $N$=1 band (${\rm N}(1520)-{\rm N}(1535)$ and
$\Delta(1620)-\Delta(1700)$).
In the spectrum of the $\Lambda$, on the other hand, it is the
negative parity flavor singlet states ( $\Lambda(1405)-
\Lambda(1520)$) that remain the lowest lying resonances,
again in agreement with experiment.
The mass splittings between the baryons with different strangeness,
and between the $\Lambda$ and the $\Sigma$
which have identical flavor, spin and flavor-spin
symmetries arise from the explicit
breaking of the $SU(3)_{\rm F}$ symmetry that is caused by the mass
splitting of the pseudoscalar meson octet, and the different
masses of the u,d and the s quarks.

This lecture has the following structure. Section 2  contains the
proof  why the commonly used perturbative gluon exchange interaction
between the constituent quarks leads to incorrect ordering of positive
and negative parity states in the spectra. In Section 3
we outline the importance of the spontaneous breaking of
chiral symmetry
for  low-energy QCD, and in Section 4 we present a
short historical sketch of the role of chiral symmetry in  quark based
models. Section 5 contains a description of the chiral boson exchange
interaction (\ref{1.1}). In Section 6 we describe the spectra of the nucleon,
the $\Delta$ resonance and the $\Lambda$ hyperon as they are predicted by
the $SU(3)_{\rm F}$ symmetric interaction (\ref{1.1}). The effect of
$SU(3)_{\rm F}$
breaking in the interaction is considered in Section 7.
Section 8 is devoted to an "exact" three-body description of baryons where
the interaction (\ref{1.1}) is taken into account to all orders.
 In Section 9
we discuss the role of the exchange current corrections to the baryon
magnetic moments that are associated with the pseudoscalar exchange
interaction. Finally, in Section 10
some recent lattice - QCD results are discussed.

\section{Why the Gluon Exchange Bears no Relation \protect\newline to 
the Baryon Spectrum}
It was accepted by many people (but not by all) that the fine splittings
in the baryon spectrum
(in analogy to atomic physics
they are often called "hyperfine splittings" as they arise
from the spin-spin forces)
are due to the gluon-exchange interaction
between the constituent quarks \cite {DERU,IGK1}. Now I shall
address myself to a formal consideration  why the
one gluon exchange interaction
cannot be relevant to the baryon spectrum.

The most important component of the one gluon exchange interaction
\cite {DERU} is the so called color-magnetic interaction

\begin{equation}H_{\rm cm}\sim-\alpha_{\rm s}\sum_{i<j} \frac {\pi}{6m_im_j}
\vec \lambda_i^{\rm C}\cdot
\vec \lambda_j^{\rm C} \vec
\sigma_i\cdot \vec \sigma_j\delta(\vec r_{ij}),\label{2.1} \end{equation}

\noindent
where  $\{\vec \lambda_i^{\rm C}\}$ are color $SU(3)$ matrices.
It is the permutational color-spin symmetry of the 3q state
which is mostly responsible for the contribution of the interaction
(\ref{2.1}). Indeed,   the corresponding two-body matrix element is

$$
<[f_{ij}]_{\rm C}\times [f_{ij}]_{\rm S}:[f_{ij}]_{\rm CS}|\vec \lambda^{\rm
C}_i \cdot \vec
\lambda_j^{\rm C}
\vec\sigma_i \cdot \vec \sigma_j|[f_{ij}]_{\rm C} \times [f_{ij}]_{\rm
S}:[f_{ij}]_{\rm CS}> $$
\begin{equation} =\left\{\begin{array}{rr}  8 &
[11]_{\rm C},[11]_{\rm S}:[2]_{\rm CS} \\  -{8\over
3} & [11]_{\rm C},[2]_{\rm S}:[11]_{\rm CS}\end{array}\right..\label{2.2}
\end{equation}

\noindent
Thus the symmetrical color-spin pairs (i.e. with a $[2]_{\rm CS}$
Young pattern) experience an attractive contribution while the
antisymmetrical ones ($[11]_{\rm CS}$) experience a repulsive contribution.
Hence the color-magnetic contribution to the $\Delta$ state ($[111]_{\rm CS}$)
is  more repulsive than to the nucleon ($[21]_{\rm CS}$) and
the $\Delta$ becomes heavier than the nucleon. The price is that $\alpha_{\rm
s}$
should be larger than unity, which is  bad. In addition there is no
empirical indication in the spectrum for a large spin-orbit component
of the gluon-exchange interaction \cite {DERU} implied by this big value of
$\alpha_{\rm s}$.

The crucial point is that the interaction (\ref{2.1}) feels only colour
and spin of the interacting quarks. Thus the structure of the ${\rm N}$ and
$\Lambda$
spectra has to be the same as these baryons differ only
by their flavour structure.
If one looks at the Particle Data Group tables, however, one immediately sees
 a different ordering of the positive and negative parity states in both
 spectra. In the ${\rm N}$ spectrum the lowest states are
${1\over 2}^+, {\rm N}(939)$; ${1\over 2}^+, {\rm N}(1440)$; ${1\over 2}^-,
{\rm N}(1535)$ $-$
${3\over 2}^-, {\rm N}(1520)$, while in the $\Lambda$ spectrum the ordering
is as follows:
${1\over 2}^+, \Lambda(1115)$;
${1\over 2}^-, \Lambda(1405)$ $-$ ${3\over 2}^-, \Lambda(1520)$;
 ${1\over 2}^+, \Lambda(1600)$.
A weak flavor dependence via the quark masses in (\ref{2.1}) cannot explain
this paradox.

The interaction (\ref{2.1}) cannot explain
why the two-quantum excitations of positive parity ${\rm N}(1440)$,
$\Delta(1600)$,
$\Lambda(1600)$ and $\Sigma(1660)$ lie below the one-quantum excitations
of negative parity  ${\rm N}(1535)$ $-$ ${\rm N}(1520)$,  $\Delta(1620) -
\Delta(1700)$,
$\Lambda(1670) - \Lambda(1690)$ and $\Sigma(1750) - \Sigma(?)$, respectively.
For instance, the positive
parity state ${\rm N}(1440)$ and the negative parity ones ${\rm N}(1535)$ $-$
${\rm N}(1520)$ have
the same mixed ($[21]_{\rm CS}$) color-spin symmetry thus the color-magnetic
contribution to these states cannot be very different (a small difference
is only due to the different radial structure of the positive and
negative parity states). But the ${\rm N}(1440)$
state belongs to the $N=2$ shell, while the ${\rm N}(1535) - {\rm N}(1520)$
pair is
a member of the $N=1$ band, which means that the ${\rm N}(1440)$ should lie
approximately $\hbar \omega$ above the ${\rm N}(1535) - {\rm N}(1520)$.
In the $\Delta$ spectrum
the situation is even more dramatic. The $\Delta(1600)$ positive parity
state has a completely antisymmetric CS-Young pattern ($[111]_{\rm CS}$),
while the negative parity states $\Delta(1620) - \Delta(1700)$ have
a mixed one. Thus
the color-magnetic contribution to the $\Delta(1600)$ is much more
repulsive than to the $\Delta(1620) - \Delta(1700)$.
In addition the $\Delta(1600)$ is the $N=2$
state, while the pair $\Delta(1620) - \Delta(1700)$ belongs to the $N=1$ band.
As a consequence the $\Delta(1600)$ must lie much higher
than the $\Delta(1620) - \Delta(1700)$. All these features are well seen
in the explicit 3-body calculations \cite{SILV}.

\section{Spontaneous Chiral Symmetry  Breaking and its Consequences
for  Low-Energy QCD}
The QCD Lagrangian with three light flavors has a global symmetry

\begin{equation}SU(3)_{\rm L} \times SU(3)_{\rm R} \times U(1)_{\rm V} \times
U(1)_{\rm A},  \label{3.1} \end{equation}

\noindent
if one neglects the masses of current u,d, and s quarks, which are small
 compared to a typical  low-energy QCD scale of 1 GeV. The $U(1)_{\rm A}$
is not a symmetry at the quantum level due to the axial anomaly. If the
$SU(3)_{\rm L} \times SU(3)_{\rm R}$ chiral symmetry of the QCD Lagrangian were
intact
in the vacuum state we would observe  degenerate
multiplets in the particle spectrum corresponding
to the above chiral group,  and all hadrons would
have their degenerate partners with opposite parity. Since this does not
happen the implication is that the chiral symmetry is spontaneously
broken down to $SU(3)_{\rm V}$ in the QCD vacuum, i.e., realized in the hidden
Nambu-Goldstone mode. A direct evidence for the spontaneously broken
chiral symmetry is a nonzero value of the quark condensates for the
light flavors

\begin{equation}<{\rm vacuum}|\bar{q}q|{\rm vacuum}> \approx -(240-250 {\rm
MeV})^3, \label{3.2} \end{equation}

\noindent
which represents the order parameter. That this is indeed so, we know from
three independent sources: current algebra \cite{GOR},
QCD sum rules \cite{SHIF}, and lattice gauge calculations \cite{DANIEL}.
There are two important generic consequences of the spontaneous chiral symmetry
 breaking. The first one is an appearance of the octet
of pseudoscalar mesons of low mass, $\pi, {\rm K}, \eta$, which represent
the associated approximate Goldstone bosons. The second one is that valence
quarks acquire a dynamical or constituent mass.
Both these
 consequences of the spontaneous chiral symmetry  breaking
are well illustrated by, e.g. the $\sigma$-model \cite{LEVY} or the Nambu and
Jona-Lasinio model \cite{Nambu}.
We cannot say at the
moment for sure what the microscopical reason for spontaneous chiral symmetry
 breaking in the QCD vacuum is. It was suggested that this occurs
when quarks propagate through instantons in the QCD vacuum \cite{SHU,DIP}.

For the low-energy baryon properties it is only essential that
beyond the spontaneous chiral symmetry breaking scale  new
dynamical degrees of freedom appear - constituent quarks and chiral
fields. The low-energy baryon properties are mainly determined
by these dynamical degrees of freedom and the confining interaction.
This is quite in contrast to pseudoscalar mesons. In the chiral limit,
$m^0_{\rm u}=m^0_{\rm d}=m^0_{\rm s}=0$,
all members of the pseudoscalar octet ($\pi, {\rm K}, \eta$) would have zero
mass,
which is most clearly seen in the Gell-Mann-Oakes-Renner \cite{GOR}
relations

$$ {m_{\pi^0}}^2 = - \frac {1}{f_\pi^2} (m_{\rm u}^0 <\overline {u}u> +
m_{\rm d}^0 <\overline {d}d>) + O({m_{\rm u,d}^0}^2), $$

$$ {m_{\pi^{+,-}}}^2 = - \frac {1}{f_\pi^2} \frac {m_{\rm u}^0+m_{\rm d}^0}
{2} (<\overline {u}u> + <\overline {d}d>) + O({m_{\rm u,d}^0}^2), $$

$$ {m_{{\rm K}^{+,-}}}^2 = - \frac {1}{f_\pi^2} \frac {m_{\rm u}^0+m_{\rm s}^0}
{2} (<\overline {u}u> + <\overline {s}s>) + O({m_{\rm u,s}^0}^2), $$

$$ {m_{{\rm K}^0,{\overline {\rm K}}^0}}^2 = - \frac {1}{f_\pi^2} \frac {m_{\rm
d}^0+m_{\rm s}^0}
{2} (<\overline {d}d> + <\overline {s}s>) + O({m_{\rm d,s}^0}^2),  $$

\begin{equation} {m_{\eta}}^2 = - \frac {1}{3f_\pi^2} (m_{\rm u}^0 <\overline
{u}u> +
m_{\rm d}^0 <\overline {d}d> + 4m_{\rm s}^0 <\overline {s}s>) + O({m_{\rm
u,d,s}^0}^2),
\label{3.3} \end{equation}

\noindent
that relate the pseudoscalar meson masses to the quark condensates
and current quark masses $m^0$. Thus the nonzero masses of mesons
are determined by the nonzero values of the current quark masses. In the
baryon case,  even in the chiral limit,
 baryons would have approximately their actual masses
of the order of 1 GeV, as these masses are mostly determined by the
dynamical (constituent) masses,  the Goldstone boson exchange interaction
among them, as well as a confining interaction.
The dynamical (constituent) masses are in turn determined mainly
 by the quark condensates, which is most clearly seen from the gap equations
of the Nambu and Jona-Lasinio model, and only weakly dependent on current
quark masses. The role of the current quark masses in baryons is
to break just the $SU(3)_{\rm F}$ symmetry in the baryon spectrum.

\section{Chiral Symmetry and the Quark Model 
\protect\newline (Historical Sketch)}
The importance of the constraints posed by chiral symmetry
for the quark bag \cite{Chthorn} and bag-like \cite{Weise}
 models for the baryons
 has been recognized early.
 In
bag models with restored chiral symmetry on the bag surface, or bag-like
models,
the
massless current quarks within the bag
were assumed to interact not only by
perturbative gluon exchange but also through chiral meson field exchange.
In these models
the chiral field  has the character of a compensating auxiliary field only,
rather than a collective low frequency Goldstone quark-antiquark
excitation. The possibility of a nonzero quark condensate
was also not addressed. As it was discussed
in a previous section, it is the quark condensate which is
the most important characteristic  determining the baryon properties.
According to the bag philosophy, however, there is a perturbative QCD
phase inside a bag, where all condensates vanish by definition.
 A general limitation of all bag and bag-like models is,
of course, the lack of translational
invariance, which is of crucial importance for a  description of the
excited states.

Common to these models is that the breaking of chiral symmetry
arises from the confining interaction.
This point of view contrasts with that of Manohar and Georgi
\cite{MAG}, who pointed out that there should
be two different scales in QCD with 3 flavors. At the first one
of these,
$\Lambda_{\chi {\rm SB}} \simeq 4 \pi f_\pi \simeq$ 1~GeV, the
spontaneous breaking of the chiral symmetry occurs, and hence
at distances
beyond $ \frac {1}{\Lambda_{\chi {\rm SB}}} \simeq$ 0.2 fm the valence
current quarks acquire their dynamical (constituent) mass
(called "chiral quarks" in \cite{MAG}), and the Goldstone bosons
(mesons) appear. The other scale, $\Lambda_{\rm QCD}
\simeq 100-300$ MeV,
is that which characterizes confinement, and the inverse of this
scale roughly coincides with the linear size of
a baryon. Between these two scales then the effective Lagrangian
should be formed out of the gluon fields that provide
a confining mechanism, as well as of the
constituent quark and pseudoscalar meson fields. Manohar and Georgi
did not, however, specify whether the baryons should be
described as bound qqq states or as chiral solitons.

The chiral symmetry breaking scale above fits well with that which
appears in the instanton liquid picture of the QCD vacuum
\cite{SHU,DIP}. In this model the quark condensates (i.e., equilibrium
of virtual quark-antiquark pairs in the vacuum state),
as well as the gluon condensate,
are supported by instanton fluctuations of a size $\sim 0.3$ fm.
Diakonov and Petrov \cite {DIP} suggested that at low momenta
(i.e., beyond the chiral symmetry breaking scale) QCD should
be approximated by an effective chiral Lagrangian of the sigma-model
type that contains valence quarks with dynamical (constituent)
masses and meson fields. They considered
a nucleon as three constituent quarks moving independently
of one another
in a self-consistent chiral field of the hedgehog
form  \cite{DIAP}.
In this picture the $\Delta$ appears
as a rotational
excitation of the hedgehog,  and no explicit confining interaction is included.
A very similar description for the nucleon was suggested in
 \cite{KAH,BIR}. These types of models are now called "quark-soliton
models"  \cite{GOE,REIN}.

The spontaneous breaking of chiral symmetry  and its consequences
-  dynamical quark mass generation,  appearance of the
quark condensate, and pseudoscalar mesons as Goldstone
excitations  - are well illustrated by the Nambu and Jona-Lasinio
model \cite{Nambu,Vogl}.
This model lacks a confining
interaction, which, as argued below, is essential for a realistic
description of the properties of  baryon physics.

\section{The Chiral Boson Exchange Interaction}
In an effective chiral  description of the baryon
structure, based on the constituent quark model, the
coupling of the quarks and the pseudoscalar Goldstone
bosons will (in the $SU(3)_{\rm F}$ symmetric approximation) have
the form $ig\bar\psi\gamma_5\vec\lambda^{\rm F}
\cdot \vec\phi\psi$ (or
$g/(2m)\bar\psi\gamma_\mu
\gamma_5\vec\lambda^{\rm F}
\cdot \psi \partial^\mu\vec\phi$),
 where $\psi$ is the fermion constituent quark
field operator, $\vec\phi$ the octet boson field
operator, and $g$ is a coupling constant. A coupling of this
form, in a nonrelativistic reduction for the constituent quark spinors,
will -- to lowest order -- give rise to a Yukawa interaction
between the constituent quarks, the spin-spin component of which has
the form
\begin{equation}V_{\rm Y} (r_{ij})=
\frac{g^2}{4\pi}\frac{1}{3}\frac{1}{4m_im_j}
\vec\sigma_i\cdot\vec\sigma_j\vec\lambda_i^{\rm F}\cdot\vec\lambda_j^{\rm F}
\{\mu^2\frac{e^{-\mu r_{ij}}}{ r_{ij}}-4\pi\delta (\vec r_{ij})\}
.\label{5.1} \end{equation}
Here $m_i$ and $m_j$ denote the masses of the interacting quarks,
and $\mu$ that of the meson. There will also be an associated
tensor component, which is discussed in ref. \cite{GLO2}.

At short range the simple form (\ref{5.1}) of the chiral boson exchange
interaction cannot be expected to be realistic and should only
be taken to be suggestive.
Because of the finite spatial extent of both the constituent
quarks and the pseudoscalar mesons
 the delta function in (\ref{5.1}) should be replaced by a finite
function, with a range of 0.6-0.7 fm, as suggested
by the
spatial extent of the mesons.
In addition, the radial behaviour of the Yukawa
potential (\ref{5.1}) is valid only if the boson field
satisfies a linear Klein-Gordon equation. The implications of the underlying
chiral symmetry of QCD
for the effective chiral Lagrangian (which in fact is not known),
which contains
constituent quarks as well as boson fields,
are that these boson fields cannot be described by linear
equations near their source. Therefore it is only at large
distances, where the amplitude of the boson fields is small, that
the quark-quark interaction reduces to the simple Yukawa form.

The latter point is rather important and has to be clarified. The
radial dependence in (\ref{5.1}) is a direct consequence of a free
$(q^2-\mu^2)^{-1}$ Green function for the boson field, and
a pseudoscalar $ig\bar\psi\gamma_5\vec\lambda^{\rm F}
\cdot \vec\phi\psi$ or pseudovector $g/(2m)\bar\psi\gamma_\mu
\gamma_5\vec\lambda^{\rm F}
\cdot \psi \partial^\mu\vec\phi$ coupling in both quark-meson vertices.
Making use of the usual "static approximation" and neglecting the
recoil corrections at both vertices in a nonrelativistic reduction for the
constituent quark spinors, one arrives at (\ref{5.1}). The free Green function
above comes from the well-known Lagrangian for a free boson field
$1/2\partial_\mu \vec\phi \partial^\mu \vec\phi - 1/2\mu^2\vec\phi^2$,
which means that the boson field satisfies a linear Klein-Gordon equation.

On the other hand, the underlying $SU(3)_{\rm R} \times SU(3)_{\rm L}$ chiral
symmetry
of QCD tells that the Goldstone boson field cannot be described by such a
simple Klein-Gordon  Lagrangian. The transformation properties of the
Goldstone boson field under $SU(3)_{\rm R}$ and  $SU(3)_{\rm L}$ chiral
rotations,
$g_{\rm R}$ and $g_{\rm L}$, are well defined if the Goldstone boson fields
$\vec\phi$
are combined in a unitary matrix \cite{COLEMAN}

\begin{equation}U(\vec\phi)=\exp{(i\frac{\vec\lambda^{\rm F} \cdot
\vec\phi}{f_\pi})}, \label{5.2} \end{equation}

\begin{equation}U' = g_{\rm R} U g_{\rm L}^{-1}. \label{5.3} \end{equation}

\noindent
The "kinetic term" in this case is

\begin{equation} \frac{f_\pi^2}{4} Tr(\partial_\mu U \partial^\mu U^{\dag} ),
\label{5.4} \end{equation}

\noindent
which in its expansion in powers of $ \vec\phi$ contains not only a
free Klein-Gordon kinetic term but also other terms of higher powers.
For instance, the terms of fourth order in the meson field give
rise to $\pi - \pi$ scattering \cite{WEIN67}, etc. The full chiral Lagrangian
describing the Goldstone boson field is unknown and should contain
a quartic term involving derivatives $\partial_\mu U$ and higher order
terms. It is clear that a dressed Green function for the field
 $ \vec\phi$ should be very different compared to the free Klein-Gordon
Green function since the selfinteraction of the Goldstone
boson field is important.
Only far away from its source (the fermion current), where the amplitude of
the boson field  $ \vec\phi$ is small, the lowest term in powers of $ \vec\phi$
becomes dominant, and therefore only at large distances the quark-quark
interaction reduces to its simple Yukawa form.

{\it At this stage the proper procedure should be to avoid further specific
assumptions about the short range behavior of
$V(r)$ in
(\ref{1.1}),  to extract instead  the required matrix elements of it
from the baryon spectrum, and to reconstruct by this an approximate
radial form of $V(r)$.
The overall minus sign in the
effective chiral boson interaction in (\ref{1.1}) corresponds to that of the
short range term in the Yukawa interaction.}

The flavor structure of the pseudoscalar octet exchange interaction
in (\ref{1.1}) between two quarks $i$ and $j$ should be understood as
follows
:
\begin{eqnarray}
V(r_{ij}) & & \vec {\lambda^{\rm F}_i} \cdot \vec {\lambda^{\rm F}_j}
\vec\sigma_i\cdot\vec\sigma_j \nonumber \\
= &  &
\left(\sum_{a=1}^3 V_{\pi}(r_{ij}) \lambda_i^a \lambda_j^a
+\sum_{a=4}^7 V_{\rm K}(r_{ij}) \lambda_i^a \lambda_j^a
+V_{\eta}(r_{ij}) \lambda_i^8 \lambda_j^8\right)
\vec\sigma_i\cdot\vec\sigma_j. \label{5.5} \end{eqnarray}
The first term in (\ref{5.5}) represents the pion-exchange interaction,
which acts only between
light quarks. The second term represents the Kaon
exchange interaction,
which takes place in u-s and d-s pair states. The $\eta$-exchange,
which is represented by the third term, is allowed
in all quark pair states. In the $SU(3)_{\rm F}$ symmetric limit
the constituent quark masses would be equal ($m_{\rm u} = m_{\rm d} = m_{\rm
s}$),
the pseudoscalar octet would be degenerate and the meson-constituent
quark coupling constant
would be flavor independent. In this limit the
form of the pseudoscalar exchange interaction reduces to (\ref{1.1}),
which does not break the $SU(3)_{\rm F}$ invariance of the baryon
spectrum. Beyond this limit the pion, Kaon and $\eta$
exchange interactions will differ ($V_\pi \not= V_{\rm K} \not= V_\eta$),
because of the difference between the strange and u, d quark
constituent masses ($m_{\rm u,d} \not= m_{\rm s}$), and because of the
mass splitting within the pseudoscalar octet
($\mu_\pi \not= \mu_{\rm K} \not= \mu_\eta$) (and possibly also because
of flavor dependence in the meson-quark coupling constant).
 The source of both the $SU(3)_{\rm F}$ symmetry
breaking constituent quark mass differences and the $SU(3)_{\rm F}$ symmetry
breaking mass splitting of the pseudoscalar octet is
the explicit chiral symmetry breaking in QCD.

\section{The Structure of the Baryon Spectrum}
The  two-quark matrix elements of the
interaction (\ref{1.1}) are:

$$<[f_{ij}]_{\rm F}\times [f_{ij}]_{\rm S} : [f_{ij}]_{\rm FS}
{}~| -V(r_{ij})\vec \lambda^{\rm F}_i \cdot \vec \lambda_j^{\rm F}
\vec\sigma_i \cdot \vec \sigma_j
{}~|~[f_{ij}]_{\rm F} \times [f_{ij}]_{\rm S} : [f_{ij}]_{\rm FS}> $$
\begin{equation}=\left\{\begin{array}{rl} -{4\over 3}V(r_{ij}) & [2]_{\rm
F},[2]_{\rm S}:[2]_{\rm FS} \\
-8V(r_{ij}) & [11]_{\rm F},[11]_{\rm S}:[2]_{\rm FS} \\
4V(r_{ij}) & [2]_{\rm F},[11]_{\rm S}:[11]_{\rm FS}\\ {8\over
3}V(r_{ij}) & [11]_{\rm F},[2]_{\rm S}:[11]_{\rm
FS}\end{array}\right..\label{6.1} \end{equation}

\noindent
{}From these the following important properties may be inferred:

(i) At short range, where $V(r_{ij})$ is positive, the chiral
interaction (\ref{1.1}) is attractive in the symmetric FS pairs and
repulsive in the antisymmetric ones. At large distances the potential
function $V(r_{ij})$ becomes negative and the situation is
reversed.

(ii) At short range,  among the FS-symmetrical pairs,
the flavor antisymmetric pairs experience
a much larger attractive interaction than the flavor-symmetric
ones, and among the FS-antisymmetric pairs
the strength of the repulsion in flavor-antisymmetric
pairs is considerably weaker than in the symmetric ones.

Given these properties we conclude, that
with the given flavor symmetry, the more symmetrical the FS
Young pattern is for a baryon the more attractive contribution at short
range comes from the interaction (\ref{1.1}). For two identical
flavor-spin Young patterns $[f]_{\rm FS}$ the attractive contribution
at short range is larger for  the more antisymmetrical
flavor Young pattern $[f]_{\rm F}$.

 Consider first, for the purposes of illustration, a schematic model
which neglects the radial dependence
of the potential function $V(r)$ in (\ref{1.1}), and assume a harmonic
confinement among quarks as well as $m_{\rm u}=m_{\rm d}=m_{\rm s}$.
In this model

\begin{equation}H_\chi\sim -\sum_{i<j}C_\chi~
\vec \lambda^{\rm F}_i \cdot \vec \lambda^{\rm F}_j\,
\vec
\sigma_i \cdot \vec \sigma_j.\label{6.2} \end{equation}

If the only interaction between the
quarks were the flavor- and spin-independent harmonic confining
interaction, the baryon spectrum would be organized in multiplets
of the symmetry group $SU(6)_{\rm FS} \times U(6)_{\rm conf}$. In this case
the baryon masses would be determined solely by the orbital structure,
and the spectrum would be organized in an {\it alternative sequence
of positive and negative parity states.}
The Hamiltonian (\ref{6.2}), within a first order perturbation theory,
 reduces the $SU(6)_{\rm FS} \times U(6)_{\rm conf}$ symmetry down to
 $SU(3)_{\rm F}\times SU(2)_{\rm S}\times U(6)_{\rm conf}$, which automatically
implies a splitting between the octet and decuplet baryons.

For the octet states ${\rm N}$, $\Lambda$, $\Sigma$,
$\Xi$ ($N=0$ shell, $N$ is the number of harmonic oscillator excitations
in a 3-quark state) as well as for their first
radial excitations of positive parity
(breathing modes) ${\rm N}(1440)$, $\Lambda(1600)$, $\Sigma(1660)$,
$\Xi(?)$ ($N=2$ shell) the flavor and spin symmetries
are $[3]_{\rm FS}[21]_{\rm F}[21]_{\rm S}$, and the contribution of the
Hamiltonian (\ref{6.2})
 is $-14C_\chi$. For the decuplet states
$\Delta$, $\Sigma(1385)$, $\Xi(1530)$, $\Omega$ ($N=0$ shell)
the flavor and spin symmetries,
as well as the corresponding matrix element, are
$[3]_{\rm FS}[3]_{\rm F}[3]_{\rm S}$
and $-4C_\chi$, respectively. The first negative parity excitations
($N=1$ shell) in the ${\rm N}$ and $\Sigma$ spectra ${\rm N}(1535)$ - ${\rm
N}(1520)$
and $\Sigma(1750)$ - $\Sigma(?)$
are described by the $[21]_{\rm FS}[21]_{\rm F}[21]_{\rm S}$ symmetries, and
the contribution
of the interaction (\ref{6.2}) in this case is $-2C_\chi$. The first negative
parity excitation in the $\Lambda$ spectrum ($N=1$ shell)
$\Lambda(1405)$ - $\Lambda(1520)$ is flavor singlet $[21]_{\rm FS}[111]_{\rm
F}[21]_{\rm S}$,
and, in this case, the corresponding matrix element is $-8C_\chi$.

These  matrix elements alone suffice to prove that
the ordering of the lowest positive and negative parity states
in the baryon spectrum will be correctly predicted by
the chiral boson exchange interaction (\ref{6.2}).
The constant $C_\chi$ may be determined from the
N$-\Delta$ splitting to be 29.3 MeV.
The oscillator
parameter $\hbar\omega$, which characterizes the
effective confining interaction,
may be determined as  one half of the mass differences between the
first excited
$\frac{1}{2}^+$ states and the ground states of the baryons,
which have the same flavor-spin, flavor and spin symmetries
(e.g. ${\rm N}(1440)$ - ${\rm N}$, $\Lambda(1600)$ - $\Lambda$, $\Sigma(1660)$
- $\Sigma$),
to be
$\hbar\omega \simeq 250$ MeV. Thus the two free parameters of this simple model
are fixed and we can  make now predictions.
In the ${\rm N}$ and $\Sigma$ sectors the mass
difference between the lowest
excited ${1\over 2}^+$ states (${\rm N}(1440)$ and $\Sigma(1660)$)
and ${1\over 2}^--{3\over 2}^-$ negative parity pairs
 (${\rm N}(1535)$ - ${\rm N}(1520)$ and $\Sigma(1750)$ - $\Sigma(?)$) will then
be
\begin{equation}{\rm N},\Sigma:\quad m({1\over 2}^+)-m({1\over 2}^--{3\over
2}^-)=250\, {\rm
MeV}-C_\chi(14-2)=-102\, {\rm MeV},\label{6.3} \end{equation}
whereas for the $\Lambda$ system ($\Lambda(1600)$,
$\Lambda(1405)$ - $\Lambda(1520)$) it should be

\begin{equation}\Lambda:\quad m({1\over 2}^+)-m({1\over 2}^--{3\over
2}^-)=250\, {\rm
MeV}-C_\chi(14-8)=74\, {\rm MeV}. \label{6.4} \end{equation}

This simple example shows how the chiral interaction (\ref{6.2})
provides different ordering of the lowest positive and negative parity excited
states in the spectra of the nucleon and
the $\Lambda$-hyperon. This is a direct
consequence of the symmetry properties of the boson-exchange interaction
discussed at the beginning of this section.
Namely, the $[3]_{\rm FS}$ state in the ${\rm N}(1440)$, $\Delta(1600)$
and
$\Sigma(1660)$ positive parity resonances from the $N=2$ band feels a
much stronger
attractive interaction than the mixed symmetry state $[21]_{\rm FS}$ in the
${\rm N}(1535)$ - ${\rm N}(1520)$,
$\Delta(1620)$ - $\Delta(1700)$
and $\Sigma(1750)$ -$\Sigma(?)$ resonances of negative parity ($N=1$ shell).
Consequently the masses of the
positive parity states ${\rm N}(1440)$, $\Delta(1600)$  and
$\Sigma(1660)$ are shifted
down relative to the other ones, which explains the reversal of
the otherwise expected "normal ordering".
The situation is different for $\Lambda(1405)$ - $\Lambda(1520)$
and
$\Lambda(1600)$, as the flavor state of  $\Lambda(1405)$ - $\Lambda(1520)$ is
totally antisymmetric. Because of this the
$\Lambda(1405)$ - $\Lambda(1520)$ gains an
attractive energy, which is
comparable to that of the $\Lambda(1600)$, and thus the ordering
suggested by the confining oscillator interaction is maintained.

Consider now, in addition, the radial dependence of the potential
with the $SU(3)_{\rm F}$ invariant version (\ref{1.1}) of
the chiral boson exchange interaction (i.e., $V_\pi (r)
=V_{\rm K} (r)=V_\eta(r)$).
If the confining interaction in each quark pair
is taken  to have the harmonic oscillator form as above,
the exact eigenvalues and eigenstates to
the coinfining 3q Hamiltonian
 are

\begin{equation}E=(N+3)\hbar\omega+3V_0,\label{6.5} \end{equation}
\begin{equation}\Psi=|N(\lambda\mu)L[f]_{\rm X}[f]_{\rm FS}[f]_{\rm F}[f]_{\rm
S}>,\label{6.6} \end{equation}

\noindent
where $N$ is the number of quanta in the state, the Elliott symbol
$(\lambda \mu)$ characterizes the $SU(3)$ harmonic oscillator symmetry,
and $L$ is the orbital angular momentum. The spatial (X), flavor-spin (FS),
flavor (F), and spin (S) permutational symmetries are indicated
by corresponding Young patterns (diagrams) $[f]$. All these functions are well
known (see, e.g.,  \cite{GLKU}).
Note that the  color state
$[111]_{\rm C}$, which is common to all the states, has been
suppressed in (\ref{6.6}). By the Pauli principle $[f]_{\rm X}=[f]_{\rm FS}$.

\begin{table}[!t]
\caption{The structure of the nucleon
and $\Delta$ resonance states up to $N=2$,
including 11 predicted unobserved or nonconfirmed
states, indicated by question marks.
The predicted energy values (in MeV) are given in the brackets
under the empirical ones.}
{\footnotesize
\begin{center}
\scalebox{.9}{
\begin{tabular}{|llll|} \hline
$N(\lambda\mu)L[f]_{\rm X}[f]_{\rm FS}[f]_{\rm F}[f]_{\rm S}$
& LS multiplet & average &$\;\;\; \delta M_\chi$\\
&&energy&\\ \hline
$0(00)0[3]_{\rm X}[3]_{\rm FS}[21]_{\rm F}[21]_{\rm S}$ & ${1\over 2}^+, {\rm
N}$ &
939&$-14 P_{00}$\\
&&&\\
$0(00)0[3]_{\rm X}[3]_{\rm FS}[3]_{\rm F}[3]_{\rm S}$ & ${3\over 2}^+, \Delta$
&
1232&$-4 P_{00}$\\
&&(input)&\\
$2(20)0[3]_{\rm X}[3]_{\rm FS}[21]_{\rm F}[21]_{\rm S}$ & ${1\over 2}^+, {\rm
N}(1440)$ &
1440&$-7 P_{00}-7P_{20}$\\
&&(input)&\\
$1(10)1[21]_{\rm X}[21]_{\rm FS}[21]_{\rm F}[21]_{\rm S}$ & ${1\over 2}^-, {\rm
N}(1535);
{3\over 2}^-, {\rm N}(1520)$ &
1527&$-7 P_{00}+ 5P_{11}$\\
&&(input)&\\
$2(20)0[3]_{\rm X}[3]_{\rm FS}[3]_{\rm F}[3]_{\rm S}$ & ${3\over 2}^+,
\Delta(1600)$ &
1600&$-2 P_{00}-2P_{20}$\\
&&(input)&\\
$1(10)1[21]_{\rm X}[21]_{\rm FS}[3]_{\rm F}[21]_{\rm S}$ & ${1\over 2}^-,
\Delta(1620);
{3\over 2}^-,\Delta(1700)$ &
1660&$-2P_{00}+6P_{11}$\\
&&(1719)&\\
$1(10)1[21]_{\rm X}[21]_{\rm FS}[21]_{\rm F}[3]_{\rm S}$ & ${1\over 2}^-, {\rm
N}(1650);
{3\over 2}^-, {\rm N}(1700)$ &
1675&$-2 P_{00}+4P_{11}$\\
&${5\over 2}^-, {\rm N}(1675)$&(1629)&\\
&&&\\
$2(20)2[3]_{\rm X}[3]_{\rm FS}[3]_{\rm F}[3]_{\rm S}$&${1\over
2}^+,\Delta(1750?);
{3\over 2}^+,\Delta(?)$&1750?&$-2P_{00}-2P_{22}$\\
&${5\over 2}^+,\Delta(?);{7\over 2}^+,\Delta(?)$&(1675)&\\
&&&\\
$2(20)2[3]_{\rm X}[3]_{\rm FS}[21]_{\rm F}[21]_{\rm S}$&${3\over 2}^+, {\rm
N}(1720);
{5\over 2}^+, {\rm N}(1680)$&1700&$-7P_{00}-7P_{22}$\\
&&(input)&\\
$2(20)0[21]_{\rm X}[21]_{\rm FS}[21]_{\rm F}[21]_{\rm S}$ & ${1\over 2}^+, {\rm
N}(1710)$
&1710&$-{7\over 2}P_{00}-{7\over 2}P_{20}+5P_{11}$\\
&&(1778)&\\
$2(20)0[21]_{\rm X}[21]_{\rm FS}[21]_{\rm F}[3]_{\rm S}$ & ${3\over 2}^+, {\rm
N}(?)
$ &?&$-P_{00}-P_{20}+4P_{11}$\\
&&(1813)&\\
$2(20)2[21]_{\rm X}[21]_{\rm FS}[21]_{\rm F}[21]_{\rm S}$ & ${3\over 2}^+, {\rm
N}(1900?);
{5\over 2}^+, {\rm N}(2000?);$ &1950?
&$-{7\over 2}P_{00}-{7\over 2}P_{22}+5P_{11}$\\
&&(1909)&\\
$2(20)2[21]_{\rm X}[21]_{\rm FS}[21]_{\rm F}[3]_{\rm S}$ & ${1\over 2}^+, {\rm
N}(?);
{3\over 2}^+, {\rm N}(?)$&1990?
&$-P_{00}-P_{22}+4P_{11}$\\
&${5\over 2}^+, {\rm N}(?);{7\over 2}^+, {\rm N}(1990?)$&(1850)&\\
&&&\\
$2(20)0[21]_{\rm X}[21]_{\rm FS}[3]_{\rm F}[21]_{\rm S}$ & ${1\over 2}^+,
\Delta(1910)
$ &1910&$-P_{00}-P_{20}+6P_{11}$\\
&&(1903)&\\
$2(20)2[21]_{\rm X}[21]_{\rm FS}[3]_{\rm F}[21]_{\rm S}$ & ${3\over 2}^+,
\Delta(1920);
{5\over 2}^+,\Delta(1905)$ &1912
&$-P_{00}-P_{22}+6P_{11}$\\
&&(1940)&\\ \hline
\end{tabular}
}
\end{center}
}
\end{table}


\begin{table}[!t]
\caption{The structure of the $\Lambda$-hyperon
states up to $N=2$, including predicted unobserved or nonconfirmed states,
indicated
by question marks. The predicted energies (in MeV)
are given in the brackets under the empirical values.}
{\footnotesize
\begin{center}
\scalebox{.9}{
\begin{tabular}{|llll|} \hline
$ N (\lambda\mu)L[f]_{\rm X}[f]_{\rm FS}[f]_{\rm F}[f]_{\rm S}$
& LS multiplet & average &$\;\;\; \delta M_\chi$\\
&&energy&\\ \hline
$0(00)0[3]_{\rm X}[3]_{\rm FS}[21]_{\rm F}[21]_{\rm S}$ & ${1\over 2}^+,
\Lambda$ &
1115&$-14 P_{00}$\\
&&&\\
$1(10)1[21]_{\rm X}[21]_{\rm FS}[111]_{\rm F}[21]_{\rm S}$ & ${1\over 2}^-,
\Lambda(1405);
{3\over 2}^-,\Lambda(1520)$ &
1462&$-12 P_{00}+4P_{11}$\\
&&(1512)&\\
$2(20)0[3]_{\rm X}[3]_{\rm FS}[21]_{\rm F}[21]_{\rm S}$ & ${1\over 2}^+,
\Lambda(1600)$ &
1600&$-7 P_{00}-7P_{20}$\\
&&(1616)&\\
$1(10)1[21]_{\rm X}[21]_{\rm FS}[21]_{\rm F}[21]_{\rm S}$ & ${1\over 2}^-,
\Lambda(1670);
{3\over 2}^-, \Lambda(1690)$ &
1680&$-7 P_{00}+5 P_{11}$\\
&&(1703)&\\
$1(10)1[21]_{\rm X}[21]_{\rm FS}[21]_{\rm F}[3]_{\rm S}$ & ${1\over 2}^-,
\Lambda(1800);
{3\over 2}^-,\Lambda(?);$ &
1815&$-2 P_{00}+4P_{11}$\\
&${5\over 2}^-,\Lambda(1830)$&(1805)&\\
&&&\\
$2(20)0[21]_{\rm X}[21]_{\rm FS}[111]_{\rm F}[21]_{\rm S}$ & ${1\over 2}^+,
\Lambda(1810)
$&1810&$-6P_{00}-6P_{20}+4P_{11}$\\
&&(1829)&\\
$2(20)2[3]_{\rm X}[3]_{\rm FS}[21]_{\rm F}[21]_{\rm S}$ & ${3\over 2}^+,
\Lambda(1890);
{5\over 2}^+,\Lambda(1820)$ &
1855&$-7 P_{00}-7P_{22}$\\
&&(1878)&\\
$2(20)0[21]_{\rm X}[21]_{\rm FS}[21]_{\rm F}[21]_{\rm S}$&${1\over
2}^+,\Lambda(?)$&
?&$-{7\over 2}P_{00}-{7\over 2}P_{20}+5P_{11}$\\
&&(1954)&\\
$2(20)0[21]_{\rm X}[21]_{\rm FS}[21]_{\rm F}[3]_{\rm S}$ & ${3\over 2}^+,
\Lambda(?)$&
?&$-P_{00}-P_{20}+4P_{11}$\\
&&(1989)&\\
$2(20)2[21]_{\rm X}[21]_{\rm FS}[21]_{\rm F}[3]_{\rm S}$ & ${1\over 2}^+,
\Lambda(?);
{3\over 2}^+,\Lambda (?);$&2020?&
$-P_{00}-P_{22}+4P_{11}$\\
&${5\over 2}^+\Lambda(?);{7\over 2}^+,\Lambda(2020?)$&(2026)&\\
&&&\\
$2(20)2[21]_{\rm X}[21]_{\rm FS}[111]_{\rm F}[21]_{\rm S}$ & ${3\over 2}^+,
\Lambda(?);
{5\over 2}^+,\Lambda(?)$&
?&$-6P_{00}-6P_{22}+4P_{11}$\\
&&(2053)&\\
$2(20)2[21]_{\rm X}[21]_{\rm FS}[21]_{\rm F}[21]_{\rm S}$ & ${3\over
2}^+,\Lambda(?);
{5\over 2}^+,\Lambda(2110)$ &2110?
&$-{7\over 2}P_{00}-{7\over 2}P_{22}+5P_{11}$\\
&&(2085)&\\ \hline
\end{tabular}
}
\end{center}
}
\end{table}

The full Hamiltonian is the sum of the confining Hamiltonian
and the chiral field interaction (\ref{1.1}).
When the
boson exchange interaction (\ref{1.1}) is treated in first order perturbation
theory, the mass of the baryon states takes the form

\begin{equation}M=M_0+N\hbar\omega+ \delta M_\chi, \label{6.7} \end{equation}

\noindent
where the chiral interaction contribution is
$ \delta M_\chi = <\Psi|H_\chi|\Psi>,$
and
$M_0 = \sum_{i=1}^3 {m_i} + 3(V_0 + \hbar \omega).$
The contribution from the chiral interaction  to each baryon
is a linear combination of the matrix elements of
the two-body potential $V(r_{12})$, defined as
\begin{equation}P_{nl}=<\varphi_{nlm}(\vec r_{12})
|V(r_{12})|\varphi_{nlm}(\vec r_{12})>.\label{6.8} \end{equation}
Here $\varphi_{nlm}(\vec r_{12})$
represents the oscillator wavefunction
with $n$ excited quanta.
 As we shall only consider
the baryon states in the $N\le 2$ bands, we shall only need
the four radial matrix elements $P_{00},P_{11},P_{20}$ and
$P_{22}$ for the numerical construction of the spectrum.

The contributions to all nucleon, $\Delta$ and $\Lambda$-hyperon states
from the boson exchange interaction, in terms of the matrix elements
$P_{nl}$, are listed in Tables 1 and 2.
In this approximate $SU(3)_{\rm F}$-invariant  version of the chiral boson
exchange interaction the $\Lambda-{\rm N}$ and the $\Xi - \Sigma$
mass differences
would solely be ascribed  to the  mass difference
between the s and u,d quarks, since all these baryons have identical
orbital structure and permutational symmetries.
The states in the $\Lambda$-spectrum would be degenerate
with the corresponding states in the $\Sigma$-spectrum which have
equal symmetries.

The  oscillator parameter $\hbar\omega$ and the four integrals
 are extracted from
the mass differences between the nucleon and the $\Delta(1232)$,
the $\Delta(1600)$ and
the ${\rm N}(1440)$, as well as the splittings between the nucleon
and the average mass of the
two pairs of states ${\rm N}(1535)-{\rm N}(1520)$ and ${\rm N}(1720)-{\rm
N}(1680)$.
This procedure yields the parameter values
$\hbar\omega$=157.4 MeV,
$P_{00}$=29.3 MeV, $P_{11}$=45.2 MeV, $P_{20}$=2.7 MeV and
$P_{22}$=--34.7 MeV. Given these values, all other excitation energies
(i.e., differences between the masses of given resonances and
the corresponding ground states)
of the nucleon, $\Delta$- and $\Lambda$-hyperon spectra are
predicted to within
 $\sim$ 15\% of the empirical values
where known, and are well within the uncertainty limits
of those values.
Note that these matrix elements provide a quantitatively satisfactory
description of the $\Lambda$-spectrum
even though they are extracted from the ${\rm N}-\Delta$ spectrum.

The relative magnitudes and signs
of the numerical parameter values can be readily understood. If
the potential function $V(\vec r)$ is assumed to have the
form of a Yukawa function with a smeared $\delta$-function
term that is positive  at short range $r\le 0.6-0.7$  fm,
as suggested by the pion size $\sqrt{<r_\pi^2>}=0.66$ fm,
one expects $P_{20}$
to be considerably smaller than $P_{00}$ and $P_{11}$,
as the radial wavefunction of the excited S-state has a node,
and as it extends further into the region  where the potential
is negative.
The negative value for $P_{22}$ is also natural, since the
corresponding wavefunction is suppressed at short range
and extends well beyond the expected
node in the potential function.


\section{The $SU(3)_{\rm F}$ Breaking Chiral Boson Interaction}
The model described above has relied on an interaction potential function
$V(r)$ in (\ref{1.1}) that is flavor independent. A refined version takes
into account the explicit flavor dependence of the potential function
in (\ref{5.5}) ($V_\pi \not= V_{\rm K} \not= V_\eta$). In the following
we show how this explicit flavor dependence provides  an
explanation of the mass splitting between the $\Lambda$ and the $\Sigma$ which
have the same quark content and the same FS, F and S
symmetries, i.e., they are degenerate within the $SU(3)_{\rm F}$ version
(\ref{1.1}) of the chiral boson exchange interaction.

Beyond the $SU(3)_{\rm F}$ limit the ground state baryons will be determined
by the $\pi$-exchange radial integral $P_{00}^\pi$, the K-exchange one,
$P_{00}^{\rm K}$, and by the $\eta$-exchange integrals,
$P_{00}^{\rm uu} = P_{00}^{\rm ud} = P_{00}^{\rm dd}$, $ P_{00}^{\rm us}$ and $
P_{00}^{\rm ss}$,
where the superscripts indicate quark pairs to which the $\eta$-exchange
applies. As indicated by the Yukawa
interaction (\ref{5.1}) these matrix elements should be inversely
proportional to the product of the quark masses of the pair state. Thus
$P_{nl}^{\rm us}={m_{\rm u}\over m_{\rm s}}P_{nl}^{\rm uu},\quad P_{nl}^{\rm
ss}=({m_{\rm u}\over
m_{\rm s}})^2P_{nl}^{\rm uu}.$
We also assume that $P_{00}^{\rm us}\simeq P_{00}^{\rm K}$, which is
suggested by the fact that the quark masses are equal in the states
in which these interactions act, and by the near equality of the Kaon
and $\eta$ masses, $\mu_\eta \simeq \mu_{\rm K}$. Thus we have only two
independent radial integrals.

 To determine
the integrals $P_{00}^\pi$, $P_{00}^{\rm K}$ and the quark mass difference
$\Delta_{\rm q}
=m_{\rm s} - m_{\rm u}$ we consider the $\Sigma(1385)-\Sigma$,
$\Delta - N$ and $\Lambda - N$ splittings:

\begin{equation}m_{\Sigma (1385)} - m_\Sigma = 4P_{00}^{\rm us} + 6P_{00}^{\rm
K}, \label{7.1} \end{equation}
\begin{equation}m_\Delta - m_{\rm N} = 12P_{00}^\pi - 2P_{00}^{\rm uu},
\label{7.2} \end{equation}
\begin{equation}m_\Lambda - m_{\rm N} = 6P_{00}^\pi - 6P_{00}^{\rm K} +
\Delta_{\rm q}, \label{7.3} \end{equation}

\noindent
which imply
$P_{00}^{\rm K}$ = 19.6 MeV,
$\Delta_{\rm q}=121$ MeV, if the conventional value of 340 MeV is given to
$m_{\rm u}$, $P_{00}^{\pi} =
28.9$ MeV and the quark mass ratio
$m_{\rm s}/m_{\rm u}=1.36$. These values of the matrix elements    lead to the
values of 65 MeV and 139 MeV for the $\Sigma-\Lambda$ and the $\Xi
-\Sigma$ mass differences

\begin{equation}m_\Sigma - m_\Lambda = 8P_{00}^\pi-4P_{00}^{\rm K} - \frac
{4}{3}P_{00}^{\rm uu}
-\frac {8}{3}P_{00}^{\rm us}, \label{7.4} \end{equation}
\begin{equation}m_\Xi - m_\Sigma = P_{00}^\pi + \frac {1}{3}P_{00}^{\rm uu}
-\frac {4}{3}P_{00}^{\rm ss} + \Delta_{\rm q} \label{7.5} \end{equation}

\noindent
  in good agreement
with the empirical values of 77 MeV and 125 MeV, respectively.

A description of the other parts of the $\Sigma$, $\Xi$ and $\Omega$ spectra
can be found in \cite{GLO2}.

\section{Three-Body Faddeev Calculations}
In the previous sections we have shown how the Goldstone boson
exchange  (GBE), taken to first order
perturbation theory and without explicit parameterizing the radial dependence,
can explain the correct level ordering of positive and negative parity states
in light and strange baryon spectra, as well as the splittings in those
spectra.
A question, however, arises about what will happen beyond  first order
perturbation theory. In order to check this we have numericaly solved
three-body
Faddeev equations \cite{GPP}. Besides the confinement potential, which is
now taken in linear form, the GBE interaction between the constituent quarks
is now included to all orders. These results further support the adequacy
of the GBE for baryon spectroscopy.

In addition to the octet-exchange interaction we include here also the
flavor-singlet ($\eta'$) exchange. In the large $N_{\rm C}$ limit the axial
anomaly becomes suppressed \cite{WITT}, and the $\eta'$ becomes
the ninth Goldstone
boson of the spontaneously broken $U(3)_{\rm L} \times U(3)_{\rm R}$ chiral
symmetry
in addition to the octet of pseudoscalar mesons \cite{COLWITT}.\\

For the GBE the spin-spin component of the interaction
between the constituent quarks $i$ and $j$ reads:

\begin{eqnarray}
V_\chi(\vec r_{ij})
&=&
\left\{\sum_{a=1}^3 V_{\pi}(\vec r_{ij}) \lambda_i^a \lambda_j^a \right.
\nonumber \\
&+& \left. \sum_{a=4}^7 V_{\rm K}(\vec r_{ij}) \lambda_i^a \lambda_j^a
+V_{\eta}(\vec r_{ij}) \lambda_i^8 \lambda_j^8
+V_{\eta^{\prime}}(\vec r_{ij}) \lambda_i^0 \lambda_j^0\right\}
\vec\sigma_i\cdot\vec\sigma_j,
\label{8.1}
\end{eqnarray}

\noindent
where $\lambda^a, a=1,...,8$ are flavor Gell-Mann matrices and
$\lambda^0 = \sqrt{2/3}{\bbbone}$.

 In the simplest case, when the boson field satisfies
the linear Klein-Gordon equation, one has the following spatial dependence
for the meson-exchange potentials in (\ref{8.1}):

\begin{equation}V_\gamma (\vec r_{ij})=
\frac{g_\gamma^2}{4\pi}\frac{1}{3}\frac{1}{4m_im_j}
\{\mu_\gamma^2\frac{e^{-\mu_\gamma r_{ij}}}{ r_{ij}}-4\pi\delta (\vec r_{ij})\}
,\label{8.2} \end{equation}

$$ (\gamma= \pi, {\rm K}, \eta, \eta'),  $$

\noindent
with quark and meson masses $m_i$ and $\mu_\gamma$, respectively.\\

Eq. (\ref{8.2}) contains both the traditional long-range
Yukawa potential as well as a
$\delta$-function term. It is the latter  that is of crucial importance
for baryon physics.
We already discussed in Section 5 that it is strictly  valid
only for pointlike particles, and that
it must be smeared out since the constituent
quarks and pseudoscalar mesons have  finite size, and
in addition the boson fields in a chiral Lagrangian
should in fact satisfy a nonlinear equation.
Furthermore it is quite natural to assume that at distances
$r \ll r_0$, where $r_0$ can be related to the
constituent quark and pseudoscalar meson sizes, there is no chiral
boson-exchange interaction, since this is the region of perturbative QCD with
the original QCD degrees of freedom. The interactions at these very
short distances are not
essential for the low-energy properties of baryons. Consequently
we use a two-parameter
"representation" for the $\delta$-function  term in (\ref{8.2})

\begin{equation}4\pi \delta(\vec r_{ij}) \Rightarrow \frac {4}{\sqrt {\pi}}
\alpha^3 \exp(-\alpha^2(r-r_0)^2). \label{8.3} \end{equation}

\noindent
Following the arguments above one should also
cut off the Yukawa part
of the GBE for $r < r_0$.

The $\pi {\rm q}$ coupling constant can be extracted from
the phenomenological pion-nucleon
coupling \cite{GLO2} as $\frac{g_8^2}{4\pi} =0.67$. For simplicity
(and to avoid any additional free parameter), the same coupling constant
is assumed for the  coupling between the $\eta$- meson and the constituent
quark.
This is  exactly in the spirit of unbroken
$SU(3)_{\rm F}$ symmetry.  For the flavor-singlet $\eta'$, however,
we must take a different coupling
$\frac{g_0^2}{4\pi}$, as the $\eta'$ decouples from the pseudoscalar
octet due to the $U(1)_{\rm A}$ anomaly.
 This fact is  illustrated best by the failure of
the Gell-Mann--Oakes--Renner relations \cite{GOR} for
the flavor singlet \cite{WE}. Lacking a
phenomenological value, we treat $g_0^2/4\pi$ as a free parameter.
The constituent masses of the u and d quarks are taken to be 340 MeV,
as suggested by the nucleon magnetic moments.

In the present calculation we neglect tensor meson-exchange forces. We expect
their
role to be of minor importance for the main features of the baryon
spectra \cite{GLO2}  (mainly due to
the absence of the strong $\delta$-function part in this case).

Our full interquark potential is thus given by

\begin{equation}V(\vec r_{ij})= V_\chi^{\rm octet}(\vec r_{ij}) +
V_\chi^{\rm singlet}(\vec r_{ij}) + Cr_{ij}. \label{8.4} \end{equation}

\noindent
While all masses and the octet coupling constant are predetermined,
we treated
$r_0$, $\alpha$, $(g_0/g_8)^2,$ and $C$ as free parameters and determined
their values to be:

$$ r_0 = 0.43 \, {\rm fm}, ~\alpha = 2.91 \, {\rm fm}^{-1},~
(g_0/g_8)^2= 1.8,~ C= 0.474 \, {\rm fm}^{-2}. $$

Notice that we do not need any constant $V_0$, which is usually
added to the confining potential. In fact only  {\it four} free
parameters  suffice to describe all 14 lowest states of the ${\rm N}$ and
$\Delta$
spectra, including the absolute value of the nucleon (ground state).
At the present stage of determining the qq potential due to GBE
we were led by the principle of working with the smallest possible
number of free parameters. Therefore we took the octet coupling constant
$g_8^2/4\pi$, and likewise the constituent quark mass $m$, as predetermined.
Of course, one may expect that subsequent studies within the GBE model
will put further constraints on the parametrization of the $qq$ potential.
\begin{figure}
\begin{center}
\epsfig{file=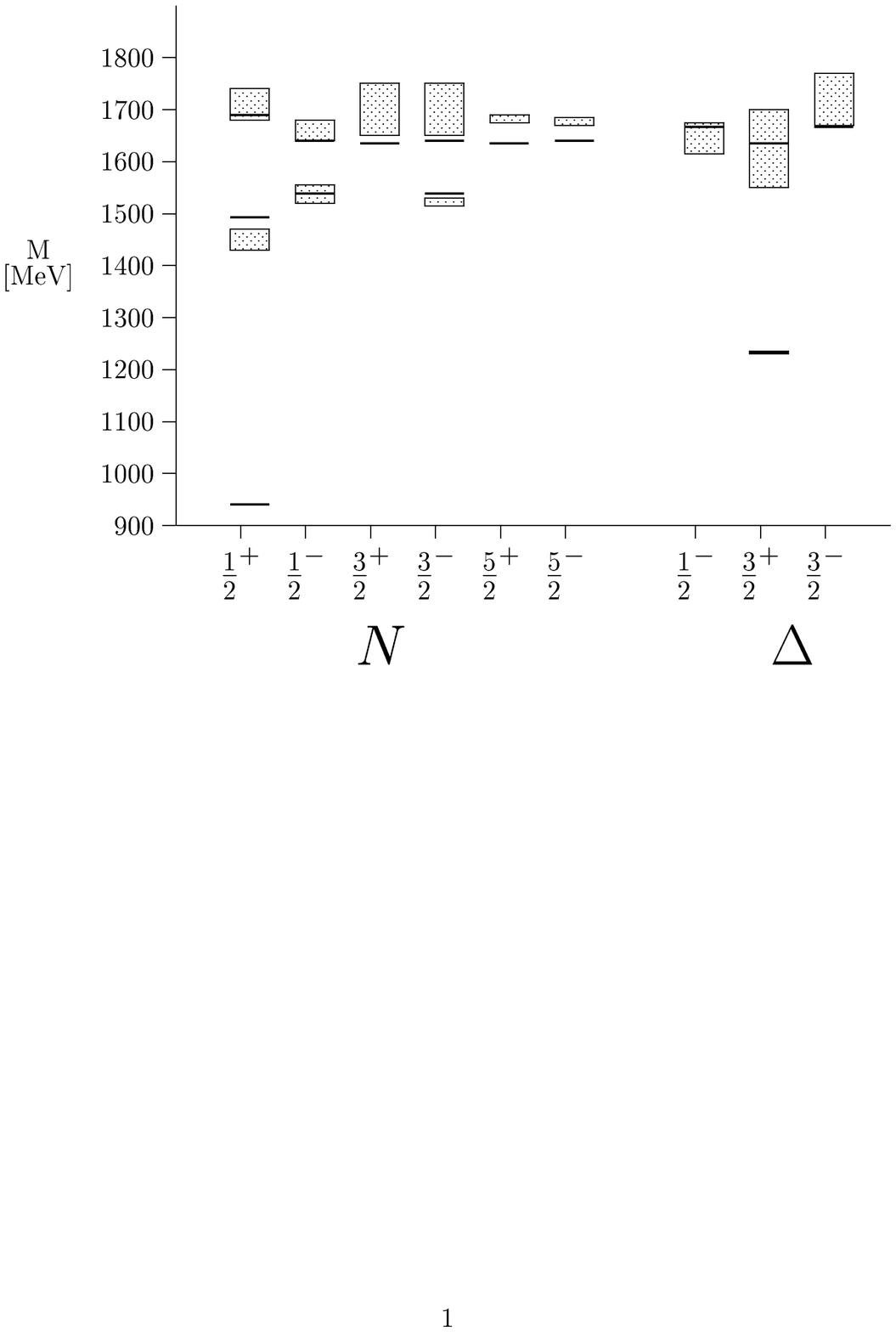, width=10cm}
\end{center}
\caption[]{Energy levels for the 14 lowest non-strange baryons with total
angular momentum and parity $J^P$. The shadowed boxes represent experimenta
uncertainties.}
\end{figure}
\begin{figure}
\begin{center}
\epsfig{file=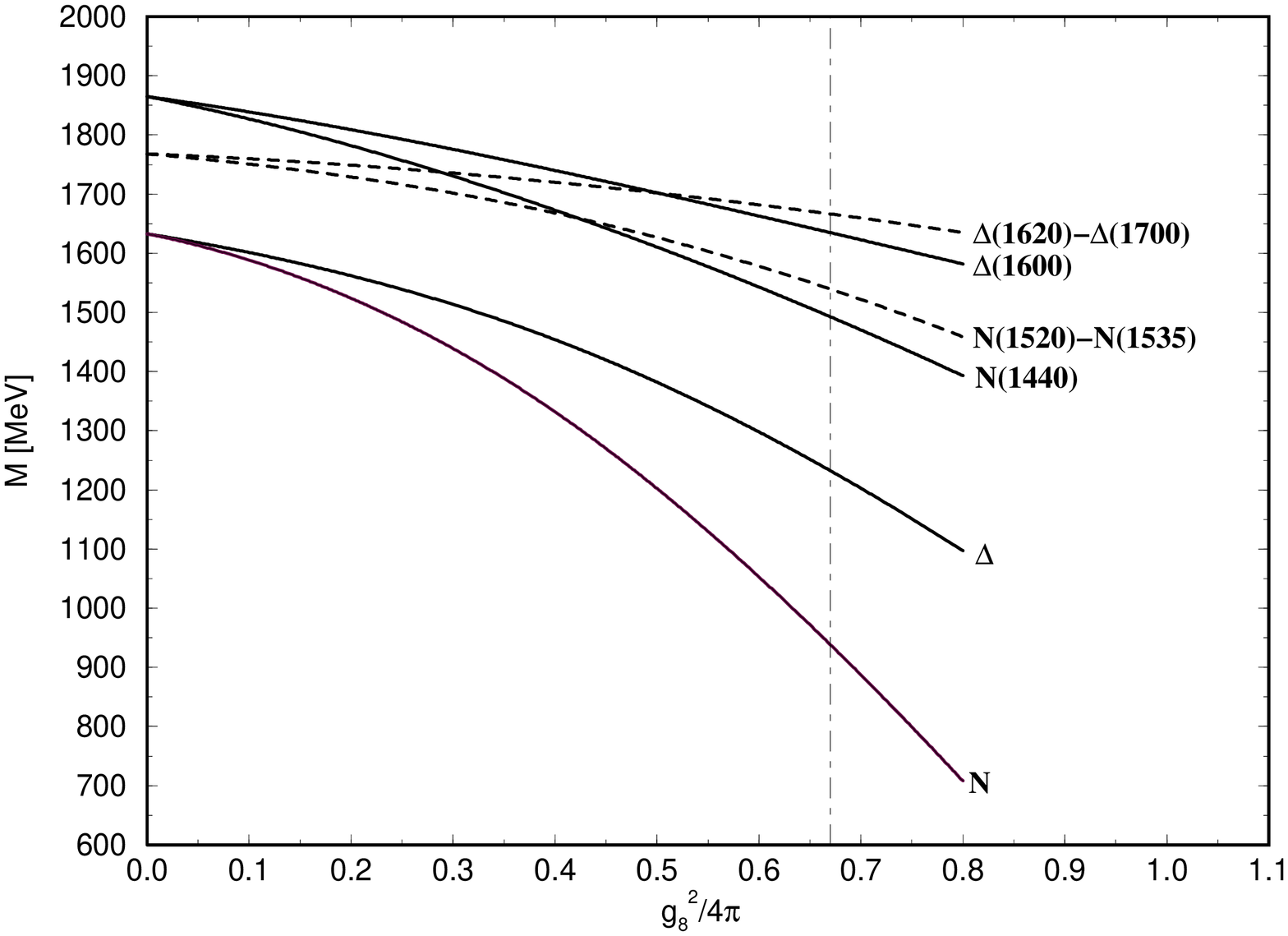, width=10cm, angle=-90}
\end{center}
\caption[]{Level shifts of some lowest baryons as a function of the strength of
the GBE. Solid and
dashed lines correspond to positive- and negative-parity states, respectively.}
\end{figure}

The qq potential (\ref{8.4}) constitutes the dynamical input into our
3-body Faddeev calculations of the baryon spectra.
We show our results  in  Fig. 1
for the  parametrization of the qq
interaction as given above.
It is well seen that the whole set of lowest ${\rm N}$ and $\Delta$
states is  reproduced quite correctly. In the most
unfavourable cases deviations  from the experimental
values do not exceed 3\%! In addition all level orderings are
correct. In particular, the positive-parity state ${\rm N}(1440)$ (Roper
resonance)
lies {\it below} the pair of negative-parity states ${\rm N}(1535)$ - ${\rm
N}(1520)$. The
same is true in the $\Delta$ spectrum with $\Delta(1600)$ and the pair
$\Delta(1620)$ - $\Delta(1700)$.
We emphasize again that
the qq potential (\ref{8.4}) is able to predict also the absolute value of the
nucleon
mass. In previous models an arbitrary constant was usually needed to achieve
the correct value of 939 MeV.

At the present stage of our investigation of the baryon spectra with the
GBE interaction we have left out the tensor forces. Therefore the
fine-structure splittings in the $LS$-multiplets are not yet introduced.
However, it is clear from the observed smallness of these splittings
and from the arguments given above, that the tensor component of the GBE
can play only a minor role. Here we also note that the Yukawa part of
the interaction in  (\ref{8.2}) is only of secondary importance. In fact, the
pattern
of Fig. 1 could also be described with the "$\delta$-part"  (\ref{8.3})
alone (and a  slightly modified set of parameters).

It is instructive to learn how the GBE affects the energy levels when
it is switched on and its strength (coupling constant) is gradually
increased (Fig. 2). Starting out from the case with confinement only,
one observes that the degeneracy of states is removed and an
inversion of the ordering of positive- and negative-parity states
is achieved, both in the ${\rm N}$ and $\Delta$ excitations. From Fig. 2 also
the
crucial importance of the chiral interaction $V_\chi$
becomes evident. Notice that
the strength of our confinement, $C$= 0.474 fm$^{-2}$, is rather small and
the confining interaction contributes much less to the splittings than
the GBE. The relative "weakness" of our effective confining interaction
could be due to a partial cancellation between the much stronger color-electric
confinement and the $\sigma$-exchange since they are of opposite sign.
Due to the same reason one cannot expect that the effective confining
interaction between the constituent quarks is strictly of linear form.\\

\section{Exchange Current Corrections \protect\newline to the Magnetic Moments}
A flavor dependent interaction of the form (\ref{1.1}) will imply the
presence of an irreducible two-body exchange current
operator, as seen, e.g., directly from the continuity equation,
by which the commutator of the interaction and the single
particle charge operator is equal to the divergence of the exchange
current density \cite{RISK}.

The general form of the octet vector exchange current operator,
that is associated
with the complete octet mediated interaction (\ref{5.5}), is
\cite{GLO2}

\begin{equation}\vec \mu^{\rm ex}=\mu_{\rm
N}\{\tilde{V}_\pi(r_{ij})(\lambda_i^1\lambda_j^2-\lambda_i^2
\lambda_j^1)
+\tilde{V}_{\rm K}(r_{ij})(\lambda_i^4\lambda_j^5-\lambda_i^5
\lambda_j^4)\}(\vec\sigma_i \times \vec \sigma_j).\label{9.1} \end{equation}

\noindent
Here $\tilde{V}_\pi(r)$ and $\tilde{V}_{\rm K}(r)$ are dimensionless
functions that describe $\pi$ and $K$ exchange, respectively,
and which include  both
 the pionic (kaonic) current and the pair current term at long range.

Consider  a simplified nonrelativistic constituent quark model.
The impulse approximation expressions for the magnetic moments of the
ground state octet baryons and their experimental values are listed in
Table 3 (columns "IA" and "exp", respectively).
A natural approach is to determine the mass ratios $m_{\rm N}/m_{\rm u}$ and
$m_{\rm N}/m_{\rm s}$ to fit the experimental values of the
magnetic moments of the
$\Sigma^-$ and $\Xi^-$ octet and   the
$\Omega$ and $\Delta^{++}$ ($\mu_{\Omega}=-2.019 \pm 0.054 ~\mu_{\rm N}$,
$\mu_{\Delta^{++}}= 4.52\pm 0.50 ~\mu_{\rm N}$ )
decuplet baryons,
which are unaffected by the exchange current operator (\ref{9.1}).
While with only two independent variables it is not possible to fit
all four experimental magnetic moments exactly, the
best overall fit,
$\mu_{\Sigma^-} = -1.00 ~\mu_{\rm N},~$ $\mu_{\Xi^-} = -0.59 ~\mu_{\rm N},~$
$\mu_{\Omega^-} = -2.01 ~\mu_{\rm N},~$
$\mu_{\Delta^{++}} = 5.52 ~\mu_{\rm N}$,
happens to be obtained with precisely the ratios
$m_{\rm N}/m_{\rm u}=2.76$ and $m_{\rm N}/m_{\rm s}=2.01$,
which were used for the constituent quark masses
to fit ground state baryons ($m_{\rm u} = 340$ MeV and $m_{\rm s} = 467$ MeV).

We find (see Table 3) that the meson exchange current contributions
systematically
improve the predictions of the naive constituent quark model (i.e., with
one-body quark currents only) for all known magnetic moments.

\begin{table}[!t]
\caption{Magnetic moments of the baryon octet (in nuclear magnetons). Column
IA contains the quark model impulse approximation expressions,
column "exp" the experimental values, column I the impulse
approximation predictions, column II the exchange current contribution
with $<\varphi_{000}(\vec r_{12})
|\tilde{V}_\pi(r_{12})|\varphi_{000}(\vec r_{12})> =  - 0.018$
and $<\varphi_{000}(\vec r_{12})
|\tilde{V}_{\rm K}(r_{12})|\varphi_{000}(\vec r_{12})> =  0.03$,
and column III the net predictions.}
{\small
\begin{center}
\begin{tabular}{|l|l|l|l|l|l|} \hline
 & IA & exp & I & II & III\\ \hline
&&&&& \\
p & ${m_{\rm N}\over m_{\rm u}}$ & +2.79 & +2.76 & +0.07 & +2.83\\
&&&&& \\
n & $-{2\over 3}{m_{\rm N}\over m_{\rm u}}$ & --1.91 & --1.84 & --0.07 &
--1.91\\
&&&&& \\
$\Lambda$ & $-{1\over 3}{m_{\rm N}\over m_{\rm s}}$ & --0.61 & --0.67 & +0.06 &
--0.61\\
&&&&& \\
$\Sigma^+$ & ${8\over 9}{m_{\rm N}\over m_{\rm u}}+{1\over 9}{m_{\rm N}\over
m_{\rm s}}$ &
+2.42 & +2.68 & --0.12 & +2.56\\
&&&&& \\
$\Sigma^0$ & ${2\over 9} {m_{\rm N}\over m_{\rm u}}+{1\over 9}{m_{\rm N}\over
m_{\rm s}}$ & ?
& +0.84 & --0.06 & +0.72\\
&&&&& \\
$\Sigma^0\rightarrow \Lambda$ & $-{1\over \sqrt{3}}{m_{\rm N}\over m_{\rm u}}$
&
$\vert$1.61$\vert$ & --1.59 & --0.01 & --1.60\\
&&&&& \\
$\Sigma^-$ & $-{4\over 9}{m_{\rm N}\over m_{\rm u}}+{1\over 9}{m_{\rm N}\over
m_{\rm s}}$ &
--1.16 & --1.00 & 0 & --1.00\\
&&&&& \\
$\Xi^0$ & $-{2\over 9}{m_{\rm N}\over m_{\rm u}}-{4\over 9}{m_{\rm N}\over
m_{\rm s}}$ &
--1.25 & --1.51 & +0.12 & --1.39\\
&&&&& \\
$\Xi^-$ & ${1\over 9}{m_{\rm N}\over m_{\rm u}}-{4\over 9}{m_{\rm N}\over
m_{\rm s}}$ & --0.65
& --0.59 & 0 & --0.59\\
&&&&& \\ \hline
\end{tabular}
\end{center}
}
\end{table}

As the constituent quarks are not too heavy, both their electromagnetic
and  axial current operators have significant relativistic correction
terms. Their effect is to reduce the magnitude of the predicted values
of both the axial coupling constants and the magnetic moments of the
baryons that are given by the static quark model. This correction reduces
the standard overprediction of the axial current coupling constant of the
nucleon (5/3 vs 1.24) and the strange baryons, but it worsens the mostly
satisfactory predictions for the magnetic moments of the baryons that are
obtained with the static quark model. In ref. \cite{DAN} it is shown
that the exchange current corrections  associated with the chiral
boson exchange interaction between the quarks can compensate for the
relativistic correction in the latter case, while leaving it operative in
the case of the axial coupling constants. This then makes it possible
to obtain  at least qualitatively satisfactory simultaneous description
of both the magnetic moments and the axial coupling constants.

\section{Instead of a Conclusion}

Instead of a conclusion we discuss  some
important recent lattice QCD results in this last section.
It was shown already
a few years ago that one can obtain a qualitatively correct
splitting between $\Delta$ and ${\rm N}$ already within a quenched
approximation (for a review and references see \cite{WEINGARTEN}).
Within the quenched approximation to QCD the sea quark closed
loop diagrams generated by gluon lines are neglected. Thus in
the quenched approximation for baryons one takes into account
only 3 continuous valence quark lines and full gluodynamics.
This quenched approximation contains, however, part of antiquark
effects related to the Z graphs formed of valence quark lines.
One can even construct diagrams within the quenched approximation
which correspond to the exchange of the color-singlet isospin 1 or 0
${\rm q}\bar{\rm q}$ pairs between valence quark lines \cite{COHEN}. It is also
important that these diagrams contribute to the baryon mass to leading
order ($\sim N_{\rm C}$) in a $1/N_{\rm C}$ expansion \cite{NC} (their
contribution
to the $\Delta - {\rm N}$ splitting appears, however, to subleading orders).

{}From the quenched measurements \cite{WEINGARTEN} it is not clear
what were
the physical reason for the $\Delta - {\rm N}$ splitting:
gluon exchanges, instantons,
or something else. To clarify this question, Liu and Dong
have recently measured the $\Delta - {\rm N}$ splitting
in the quenched and a further so-called
"valence approximation" \cite{LIU}. In the valence approximation the
quarks are limited to propagating only forward in time (i.e., Z graphs
and related quark-antiquark pairs are removed). The gluon exchange and all
other possible gluon configurations, including instantons, are exactly the
same in both approximations. The striking result is that the $\Delta - {\rm N}$
splitting is observed only in the quenched approximation but not in the valence
approximation, in which the ${\rm N}$ and the $\Delta$ levels are degenerate
within
error bars. Consequently the $\Delta - {\rm N}$ splitting must receive a
considerable
contribution from  the diagrams with ${\rm q}\bar{\rm q}$ excitations, which
correspond
to the meson exchanges, but not from the gluon exchange or instanton-induced
interaction between quarks (to be precise, the instanton-induced interaction
could be rather important for the interactions between quarks and antiquarks).

If the observation of Liu and Dong is confirmed
it would be important
to measure the relative positions of the lowest excited states of
positive and negative parity in the
${\rm N}$, $\Delta$, $\Lambda$ and $\Sigma$ spectra
within both the quenched and the valence approximation. One expects that,
if the entire $\Delta-{\rm N}$ splitting (i.e., 300 MeV) is due to the
antiquark
excitations in the quenched approximation, then $1/2^+,{\rm N}(1440)$ should be
below
the negative parity pair $1/2^-,{\rm N}(1535) - 3/2^-,{\rm N}(1520)$,
while in the $\Lambda$
spectrum the situation should be opposite: the negative parity pair
$1/2^-,\Lambda(1405) - 3/2^-,\Lambda(1520)$
should be below the first positive
parity excitation $1/2^+,\Lambda(1600)$. In the valence approximation the
spin-spin force among quarks, which is due to Goldstone boson exchange,
is absent, and the relative position of $1/2^+,{\rm N}(1440)$ and
$1/2^-,{\rm N}(1535) - 3/2^-,{\rm N}(1520)$
should be just opposite  to the quenched approximation: the
$1/2^+,{\rm N}(1440)$ should be above the negative parity pair. However,
in the $\Lambda$ spectrum the negative parity pair
$1/2^-,\Lambda(1405) - 3/2^-,\Lambda(1520)$
should  still be below the $1/2^+,\Lambda(1600)$.
\vspace{1.0cm}

\noindent {\bf  Acknowledgements}

\noindent It is a pleasure to thank Dan Riska, Zoltan Papp and Willi Plessas
for their collaboration that was crucial for the results presented in this
lecture.

\end{document}